\documentclass[twocolumn,showpacs,preprintnumbers,amsmath,amssymb]{revtex4}
\usepackage{amsfonts}

\usepackage{epsf}
\usepackage{anysize}
\usepackage{hyperref}
\usepackage{graphicx}
\usepackage{dcolumn}
\usepackage{bm}
\usepackage{amsmath}
\usepackage{amssymb}
\usepackage{mathrsfs}
\usepackage[latin1]{inputenc}
\usepackage{makecell,multirow,diagbox}

\makeatletter
  \newcommand\figcaption{\def\@captype{figure}\caption}
  \newcommand\tabcaption{\def\@captype{table}\caption}
\makeatother

\begin{document}


\title{Learning Robust and High-Precision Quantum Controls}

\author{Re-Bing Wu and Haijin Ding}
\affiliation{Department of Automation, Tsinghua University,
Beijing, 100084, P.R. China \\ Center for Quantum Information Science and Technology, BNRist,
Beijing, 100084, P.R. China}\email{rbwu@tsinghua.edu.cn}
\author{Daoyi Dong}
\affiliation{School of Engineering and
Information Technology, University of New South Wales, Canberra, ACT
2600, Australia}\email{daoyidong@gmail.com}
\author{Xiaoting Wang}
\email{xiaoting@uestc.edu.cn}
\affiliation{Institute of Fundamental and Frontier Sciences, University of Electronic Science and Technology of China, Chengdu,610054, China}

\date{\today}

\begin{abstract}
Robust and high-precision quantum control is extremely important but challenging for the functionization of scalable quantum computation. In this paper, we show that this hard problem can be translated to a supervised machine learning task by treating the time-ordered quantum evolution as a layer-ordered neural network (NN). The seeking of robust quantum controls is then equivalent to training a highly {\it generalizable} NN, to which numerous tuning skills matured in machine learning can be transferred. This opens up a door through which a family of robust control algorithms can be developed. We exemplify such potential by introducing the commonly used trick of batch-based optimization, and the resulting stochastic b-GRAPE algorithm is numerically shown to be able to remarkably enhance the control robustness while maintaining high fidelity.
\end{abstract}

\maketitle
%
\section{Introduction}
Highly accurate and stable control of quantum hardware is crucial for achieving expected quantum supremacy in the near future \cite{Harrow2017}. Usually, the control design is easy with respect to a deterministic model. However, finding a single-shot solution that also tolerates system's uncertainties, e.g., imprecisely identified parameters \cite{Dahleh1990} or time-varying noises in the Hamiltonian \cite{Demiralp1998}, is much harder. In the literature, this problem has been tackled from various aspects. Most of these evaluate the control robustness by the geometric curvature of some high-dimensional manifold \cite{Hocker2014}, which can minimized to enhance the robustness. This point of view leads to various expansion-based methods that have been experimentally very successful, including the adiabatic approach (STIRAP) \cite{Vitanov2017} against control pulse imprecisions, dynamical decoupling (DD) \cite{Viola1999,Biercuk2009,Souza2011,Green2013,Santos2008} and difference evolution (SuSSADE) \cite{Zahedinejad2015,Zahedinejad2016} algorithms against environmental noises, and other Taylor-expansion based approaches \cite{Daems2013,Huang2017}.

From a different but more unified point of view, the control against uncertainties can be thought as manipulating a collection of quantum systems under a uniform control. Along this route, ensemble- or sampling-based approaches \cite{Li2006,Ruths2012,Chen2014} were proposed to minimize the average error of the entire collection or a subset of samples of them, respectively. These algorithms have been successful in overcoming inhomogeneity of control fields in NMR experiments, and are in principle applicable to arbitrary type of uncertainties, which can vary with time or not. Compared to the geometric approaches, sampling-based methods are not restricted to the perturbation regime and can thus explore larger uncertainties. However, in practice they are limited to low-dimensional systems or systems with few uncertainty parameters due to the exponentially increasing computational cost. Even if the computation is affordable, the search for robust controls is often hindered by poor solutions due to the loss of controllability over a large collection of sampled control quantum systems.

Our studies follow the latter route for its applicability to non-perturbation regimes and capability of dealing with different types of uncertainties. We find that the search for robust quantum controls can be formulated as a supervised learning task, and the controlled quantum evolution is taken as a deep neural network (DNN) to be trained for accomplishing the task. More importantly, the pursuit of control robustness can be naturally translated to the training goal of a highly {\it generality} DNN model \cite{Xu2012}. This connection provides a new angle for understanding and solving robust quantum control problems enlightened by vast studies in deep learning (DL). For examples, our algorithm to be presented in this paper is illuminated by the ways of improving the generalizability of a DNN \cite{Bottou2018}, from the following two aspects:

(1) {\it Data augmentation.} One always learns better with more samples. Many DL problems have to learn from limited labeled training samples that are hard to obtain (e.g., diagnosis of professional doctors from medical images). However, as will be shown later, unlimited number of training samples are available for the training in our problem setting,

(2) {\it Mini-batch optimization.} In classical big-data applications, a smart way of alleviating the computational burden is to, instead of evaluating the loss with all samples, calculate the loss and gradient functions with randomly selected mini-batches of samples that vary from iteration to iteration. In this way, an unlimited number of samples can be explored after sufficiently many iterations. More importantly, the noisy and thus less stable training dynamics can effectively improve the {\it generalizability} by pulling the search away from the weakly attractive (i.e., non-robust) solutions. This merit has been extensively approved in the practice of DL.

In the following, we will show how the robust quantum control problem is translated to a supervised learning task, and how the mini-batch training skill is employed to improving the robustness of quantum controls. The rest of this paper is organized as follows. Section \ref{Sec:b-GRAPE} presents the b-GRAPE algorithm, following which Section \ref{Sec:Simulation} demonstrates the effectiveness of b-GRAPE algorithm via two typical examples that involve time-invariant parametric uncertainties and time-varying noises, respectively. Finally, conclusion is drawn in Section \ref{Sec:Conclusion}.

\section{Mini-batch training of robust and high-precision quantum controls}\label{Sec:b-GRAPE}
In this section, we first show how the robust control design problem can be translated to the training of a generalizable learning model, following which the b-GRAPE algorithm is presented by incorporating the mini-batch training into the GRAPE optimization process.
\subsection{Robust control design as a supervised learning task}
Let us start from the general model of uncertain quantum control systems:
\begin{eqnarray}\label{eq:Schoedinger equaiton}
\dot{U}(t,\epsilon) & = & -iH\left[u(t),\epsilon\right]U(t,\epsilon),
\end{eqnarray}
in which the $N\times N$ unitary propagator $U(t,\epsilon)$ is steered from the identity matrix $\mathbb{I}_N$ by the control function $u(t)$. The variable $\epsilon\in\mathbb{R}^k$ denotes the uncertainties in the Hamiltonian that can be some constant but unknown parameters or time-dependent noises (discretized into a vector of uncertainty parameters). We expect to find a robust control $u(t)$ that steers the gate operation $U(T,\epsilon)\in \mathbb{C}^{N\times N}$ to the target gate $U_f$ for all possible $\epsilon$. Since such goal is usually not achievable, we can approach it by minimizing the average infidelity
\begin{equation}\label{eq:population risk}
  L[u(t)]=\int_{\mathbb{R}^k}\|U(T,\epsilon)-U_f\|^2P(\epsilon){\rm d}\epsilon,
\end{equation}
where $P(\epsilon)$ is the {\it a priori} probability distribution of the uncertainty parameter. To alleviate the computation burden of the integral (\ref{eq:population risk}), we approximate the average infidelity over a finite number of uncertainty parameters sampled from $P(\epsilon)$, say $\mathcal{S}=\{\epsilon_1,\epsilon_2,\cdots\}$, as follows
\begin{equation}\label{Eq:L_S}
 L[u(t),\mathcal{S}] = |S|^{-1}\sum_{\epsilon\in S} \|U(T,\epsilon)-U_f\|^2.
\end{equation}
The so-called ensemble-based and sample-based algorithms are subject to the above two cost functions (\ref{eq:population risk}) and (\ref{Eq:L_S}), respectively.

\begin{figure}
\centering
\includegraphics[width=1\columnwidth]{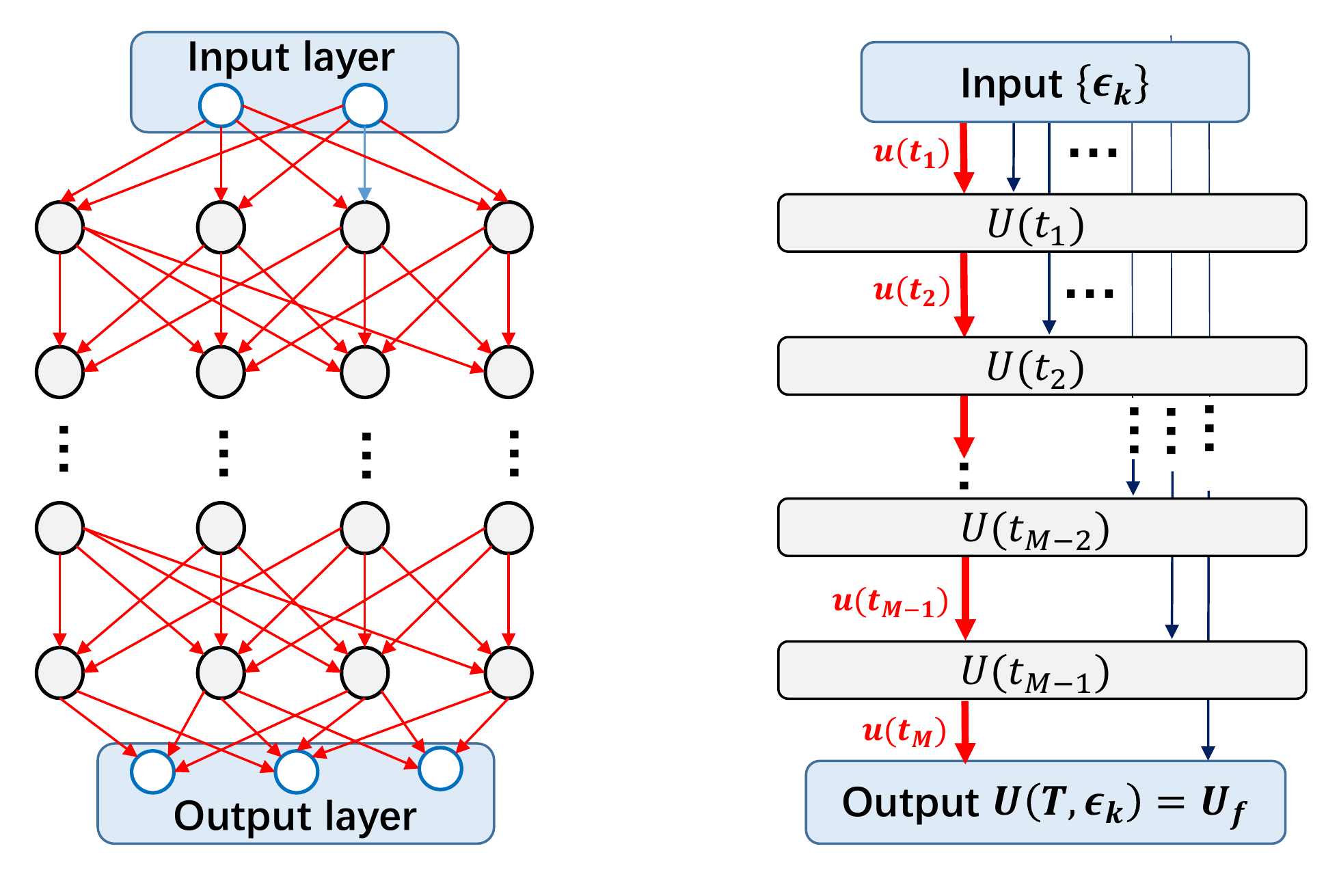}
\caption{The similarity between a deep neural network (left) and a quantum system under piecewise controls (right). The quantum system outputs a unitary propagator from a given input $\epsilon$, where the time-ordered control amplitudes $u(t_1),u(t_2),\cdots,u(t_M)$ play the role of hyper-parameters in a quantum neural network.}
\label{fig:DNN}
\end{figure}

The way we improve the robustness through minimizing the average loss of fidelity is actually the same as a supervised machine learning process through minimizing its empirical risk on a set of training samples. As is illustrated in Fig.~\ref{fig:DNN}, the quantum control system can be envisioned as a linear neural network (NN) that outputs a unitary propagator from an input uncertainty parameter $\epsilon$. Under piecewise-constant controls (e.g., generated from an Arbitrary Waveform Generator), the unitary propagator at each sampling time corresponds to a layer in an NN (with width being $N^2$), while the time-ordered control amplitudes play the role of weight parameters between adjacent layers. Note that the equivalent quantum neural network does not have a standard feed-forward structure, because every layer is affected by the network's input (i.e., the uncertainty parameter $\epsilon$). In this regard, the network is more like an residue neural network (ResNet) \cite{He2015} in which shortcuts can be made between non-adjacent layers. In this way, we may translate the robust quantum control design to a supervised deep learning (DL) task that aims at finding an NN model that outputs the same desired quantum gate $U_f$ for all input uncertainty samples. It should be noted that this picture is different from recent robust quantum control studies that are also inspired by machine learning \cite{Yang2018,Bukov2018,Arrazola2018,YuezhenNiu2018}. These existing works introduce external {\it artificial} DNNs to the training of robust quantum controls, but we take the controlled quantum system itself as a {\it natural} quantum DNN.

\subsection{b-GRAPE algorithm}

Now let us see how the mini-batch technique is applied to the search for robust quantum controls. We can randomly sample the uncertainty parameters $\epsilon_k$, and each pair $(\epsilon_k,U_f)$ forms a labeled sample for the supervised learning. Note that we have {\it uncountably} many labeled samples [i.e., pairs of $\epsilon\rightarrow U_f$ for all admissible $\epsilon\in\mathbb{R}^k$] that can be used for free, although they cannot be efficiently exploited by the existing sampling-based algorithms due to the required computing resource with large number of samples.

Concretely, we first choose a proper batch size, say $B$, and sample the batches according to the probability distribution $P(\epsilon)$, forming
\begin{equation}\label{}
  \mathcal{S}^{(j)}=\{\epsilon_1^{(j)},\cdots,\epsilon_B^{(j)}\},
\end{equation}
where $\quad j=1,2,\cdots$ are the indices of iterations. These batches are used for calculating the gradient direction in each iteration:
\begin{eqnarray}
  g[u(t),\mathcal{S}^{(j)}] &=& \frac{\delta L[u(t),{\mathcal{S}^{(j)}}]}{\delta u(t)} \nonumber \\
    &=& \frac{1}{B}\sum_{k=1}^B\frac{\delta \|U(T,\epsilon_k^{(j)})-U_f\|}{\delta u(t)}. \label{eq:g(t)}
\end{eqnarray}

The simplest control updating strategy is to take the ``steepest descending" direction along the stochastic gradient (\ref{eq:g(t)}) with some prescribed learning rate $\alpha_j$:
\begin{equation}\label{Eq:sgd}
  u^{(j+1)}(t) = u^{(j)}(t) -\alpha_j\cdot g[u^{(j)}(t),\mathcal{S}^{(j)}].
\end{equation}
However, the noisy gradient (\ref{eq:g(t)}) caused by randomly chosen batches may distabilize the steepest-descent iteration, especially when the batch size is very small. To stabilize the training dynamics, one can introduce a momentum term, i.e., the gradient direction in the previous iteration, to reduce the variance of the loss function:
\begin{equation}\label{Eq:sgdm}
\begin{split}
  u^{(j+1)}(t) = &    u^{(j)}(t) - \alpha_j\cdot \left\{\lambda g[u^{(j)}(t),\mathcal{S}^{(j)}]\right. \\
    &   \left.+(1-\lambda) g[u^{(j-1)}(t),\mathcal{S}^{(j-1)}]\right\}.
\end{split}
\end{equation}
In practice, the weight parameter $\lambda$ is usually chosen as a small positive real number (e.g., 0.1 or 0.01), so that the iteration is dominated by the momentum.

For simplicity, we term the proposed algorithm as b-GRAPE (``b" stands for ``batch''), which can be easily remoulded from the renowned GRAPE algorithm \cite{Khaneja2005} that has been extensively applied for quantum control. Correspondingly, we denote the sample-based algorithm as s-GRAPE for (``s" for ``sampling") \cite{Chen2014}. The s-GRAPE algorithm is actually a special case of b-GRAPE algorithm when using a fixed batch in all iterations, while GRAPE (for deterministic quantum systems) is a special case of s-GRAPE when only one sample is used.

Note that Our algorithm is distinct from SPSA (Simultaneous Perturbation Stochastic Algorithm), another type of stochastic gradient algorithm,  which calculates the gradient by randomizing the projected direction instead of the samples. The latter had been proposed for online model-free learning of robust quantum control and tomography ~\cite{Ferrie2015}, and it can be combined with b-GRAPE for broader applications.

\section{Numerical Simulations}\label{Sec:Simulation}
In this section, we show by two simulation examples how the DL-illuminated b-GRAPE algorithm can effectively harden quantum controls by learning from the uncertainties.

\subsection{Example 1: parametric uncertainty}
The first example considers time-invariant parametric uncertainties in a three-qubit control system:
\begin{equation}\label{}
\begin{split}
   H(t)& = (1+\epsilon_1)\sigma_{1z}\sigma_{2z}+(1+\epsilon_2)\sigma_{2z}\sigma_{3z} \\
    & +\sum_{k=1}^3\left[u_{kx}(t)\sigma_{kx}+u_{ky}(t)\sigma_{ky}\right],
\end{split}\end{equation}
where $\sigma_{k\alpha}$, $k=1,2,3$ and $\alpha =x,y,z$, are the Pauli operators on the $k$-th qubit. The uncertainty parameters $\epsilon_1$ and $\epsilon_2$ represent the identification errors in the coupling constants (dimensionless after normalization). Each qubit is manipulated by two independent control fields $u_{kx}(t)$ (along $x$-axis) and $u_{ky}(t)$ (along $y$-axis), respectively. The target three-qubit gate $U_f$ is chosen as the Toffoli gate (or controlled-controlled-NOT gate).

To start with, we set the time duration as $T=10$ and divide each control field evenly into $M=100$ piecewise constant segments. Assume that the uncertain coupling constants vary by at most $\pm 20\%$, we uniformly sample $\epsilon_1$ and $\epsilon_2$ from the set $\mathcal{S}=\{(\epsilon_1,\epsilon_2):|\epsilon_{1}|\leq 0.2,|\epsilon_2|\leq0.2\}$. The b-GRAPE algorithm is tested under three typical batch sizes $B=1$, $10$ and $100$, and is compared with s-GRAPE under identical batch sizes and initial guesses on the control. Because the training process is more stable under large batches, the learning rates are corresponding chosen as $\alpha=0.002$, $0.02$ and $0.2$, respectively, to be proportional to the batch size.

\begin{figure}
\centering
\includegraphics[width=1\columnwidth]{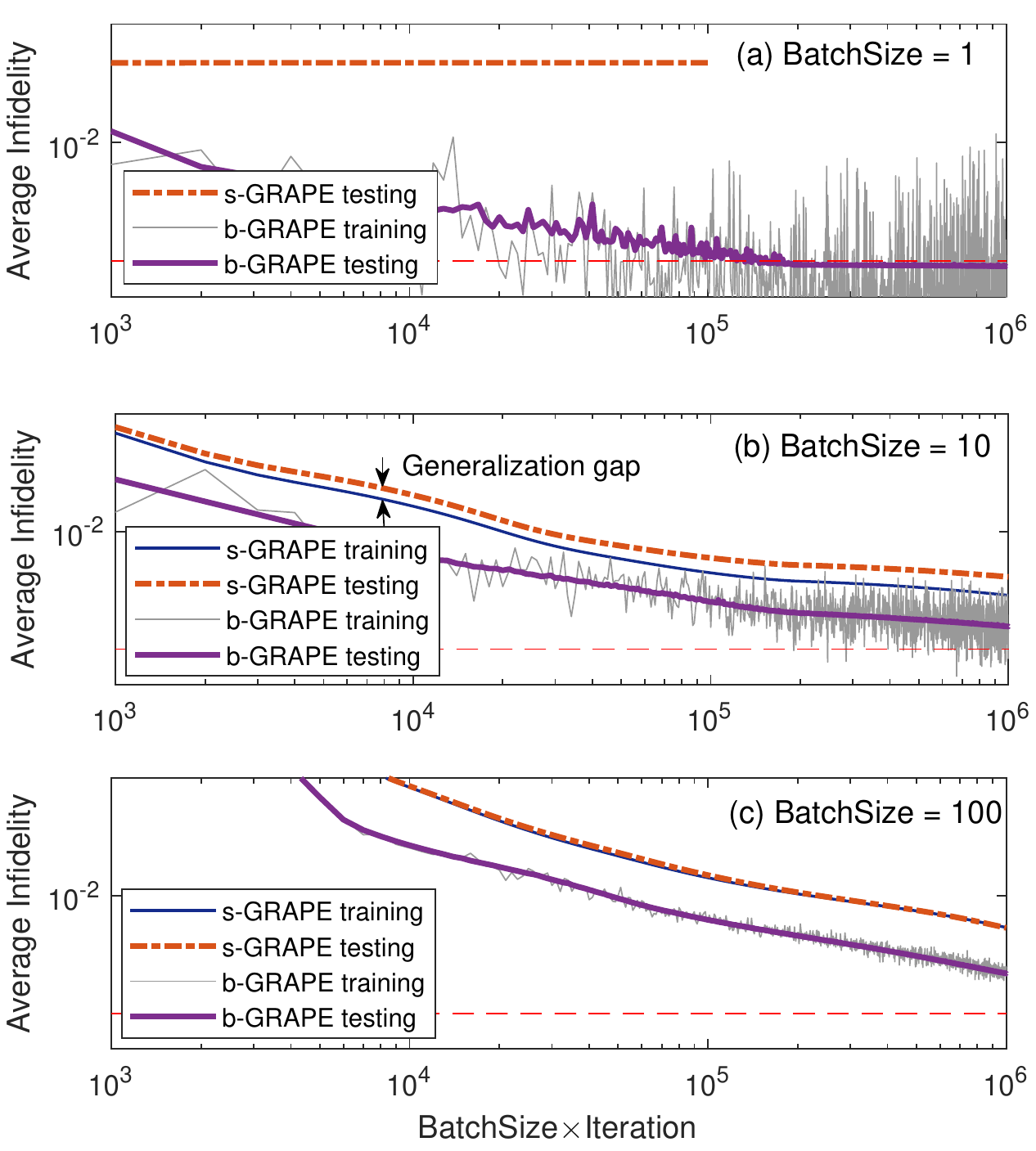}
\caption{The training and testing curves with b-GRAPE and s-GRAPE algorithms, where the control time is $T=10$, the number of control segments is $M=100$, and the batch sizes are(a) $B=1$, (b) $B=10$ and (c) $B=100$. In all cases, the batch-based b-GRAPE algorithm outperforms the sample-based s-GRAPE algorithm.}
\label{fig:learningcurve}
\end{figure}

In the simulations, we optimize the control fields along the momentum-based stochastic gradient (\ref{Eq:sgdm}). The resulting training curves, namely the average infidelity evaluated on each batch versus the number of evaluated samples (equal to the batch size times the number of iterations), are shown in Fig.~\ref{fig:learningcurve} for both b-GRAPE and s-GRAPE algorithms. The batch-induced noises can be seen in all b-GRAPE training curves, whose variance is large when using small batches. Nevertheless, an evident trend of decrease is still observable.

Because the average infidelity calculated with small batches may not reflect the actual performance, we reevaluate the performance of the control functions obtained in each iteration by 1000 independent testing samples drawn from the same probability distribution, as a better approximation of the true average infidelity (\ref{eq:population risk}). For s-GRAPE with batch size $B=1$, the training error can approach the computer machine precision, which is far below the range displayed in the plot, but the testing performance is very poor ($\approx 0.1$). Such {\it overfitting} characteristic is more clearly indicated by the gap between the testing curve (above) and the training curve (below) when $B=10$. In contrast, the testing curves of b-GRAPE always fit (on average) very well to the training curves, exhibiting much better generalizability owing to the ability of exploring many more samples.

The most significant difference, as can be seen with all tested batch sizes, is that b-GRAPE finds much more robust controls than s-GRAPE. For the example of $B=100$ [see Fig.~\ref{fig:learningcurve}(c)], the generalization gap of s-GRAPE is almost invisible, implying that the batch size has been sufficiently large to avoid overfitting. However, b-GRAPE still performs much better than s-GRAPE, owing to the batch-induced noises that steer the search away from poorer solutions. The best result in all simulations is achieved when using the smallest batch size $B=1$ [see Fig.~\ref{fig:learningcurve}(a)], under which the average infidelity can be reduced to be below 0.001 (lower than the error correction threshold). The contrast again testifies for the active role of batch-induced noises, which is the strongest when $B=1$, on guiding the search toward more robust solutions.

\begin{figure}
\centering
\includegraphics[width=1.0\columnwidth]{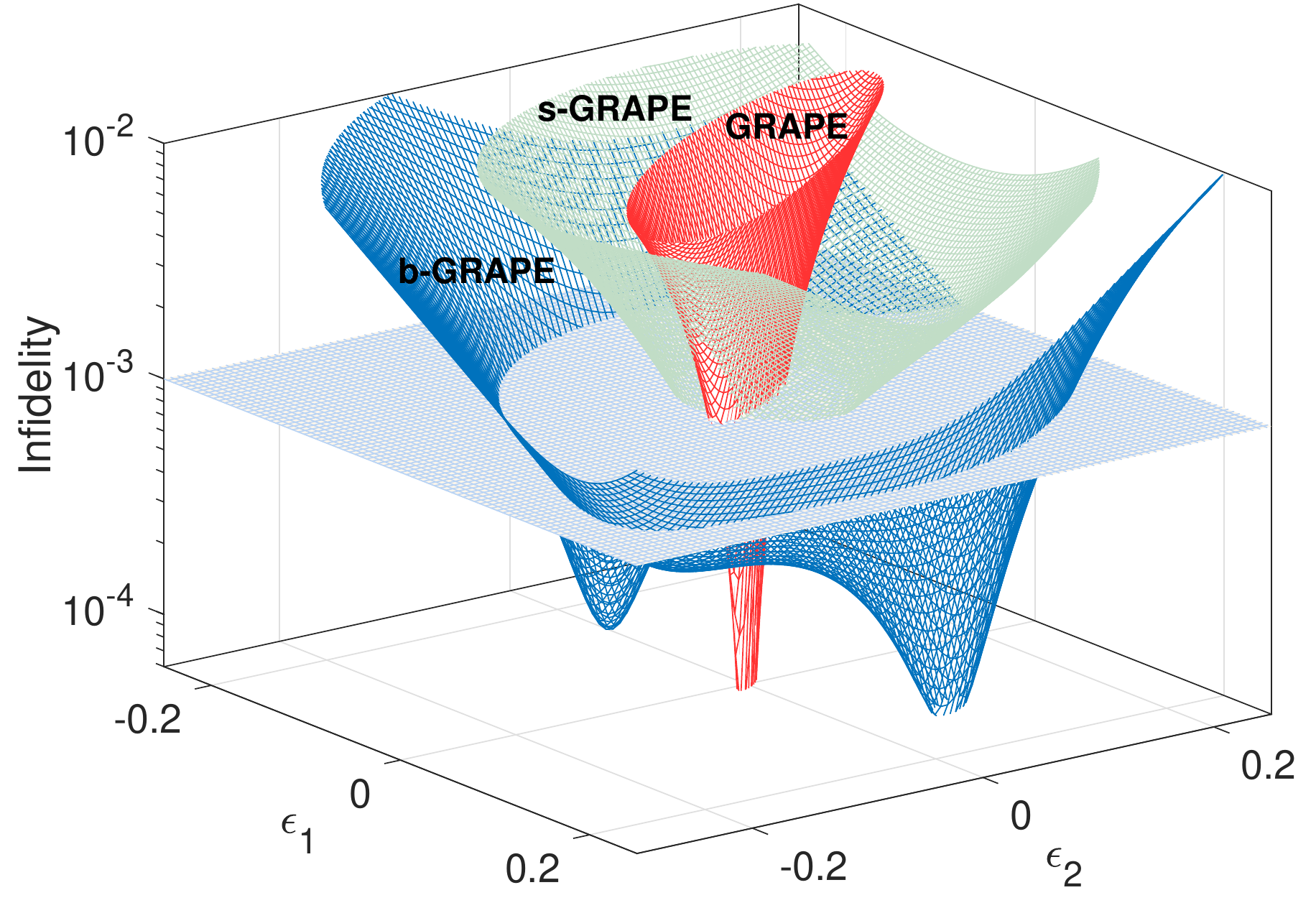}
\caption{The robustness control landscape defined as the gate infidelity versus two uncertainty parameters under controls optimized with GRAPE (red), s-GRAPE (white) using 100 samples and b-GRAPE with batch size $B=1$ (blue). The b-GRAPE obtains a much more robust control than the other two schemes, where the robustness is quantified by the area enclosed by the level set at 0.001. }
\label{fig:robust3D}
\end{figure}

To manifest the degree of robustness enhancement, we compare control fields obtained by GRAPE (fixed sample $\epsilon_1=\epsilon_2=0$), s-GRAPE ($B=100$, fixed and large batch) and b-GRAPE ($B=1$, random mini-batch) algorithms via their 3-D robustness landscapes (i.e., the infidelity versus the two uncertainty parameters, see Fig.~\ref{fig:robust3D}). The landscape also facilitates the quantification of control robustness, which can be evaluated by the area enclosed by the level set at some threshold value (say 0.001 in the figure, which is below the quantum error correction threshold \cite{Bennett1996}). The control obtained by GRAPE achieves extremely high precision at the chosen sample $\epsilon_1=\epsilon_2=0$, but it is very sensitive to the uncertainty as indicated by the sharp minimum. By contrast, the landscapes corresponding to s-GRAPE and b-GRAPE algorithms are much flatter. The obtained controls maintain high precision in a much broader region, at the price of sacrificing the precision at the center. Quantitatively, s-GRAPE enhances the robustness by about 4 times, and the control found by b-GRAPE is more than 10 time stronger than that of s-GRAPE. The level set at $0.001$ achieved by b-GRAPE almost fills the full $0.2\times 0.2$ square from which the training samples are drawn.

\begin{figure}
\centering
\includegraphics[width=1\columnwidth]{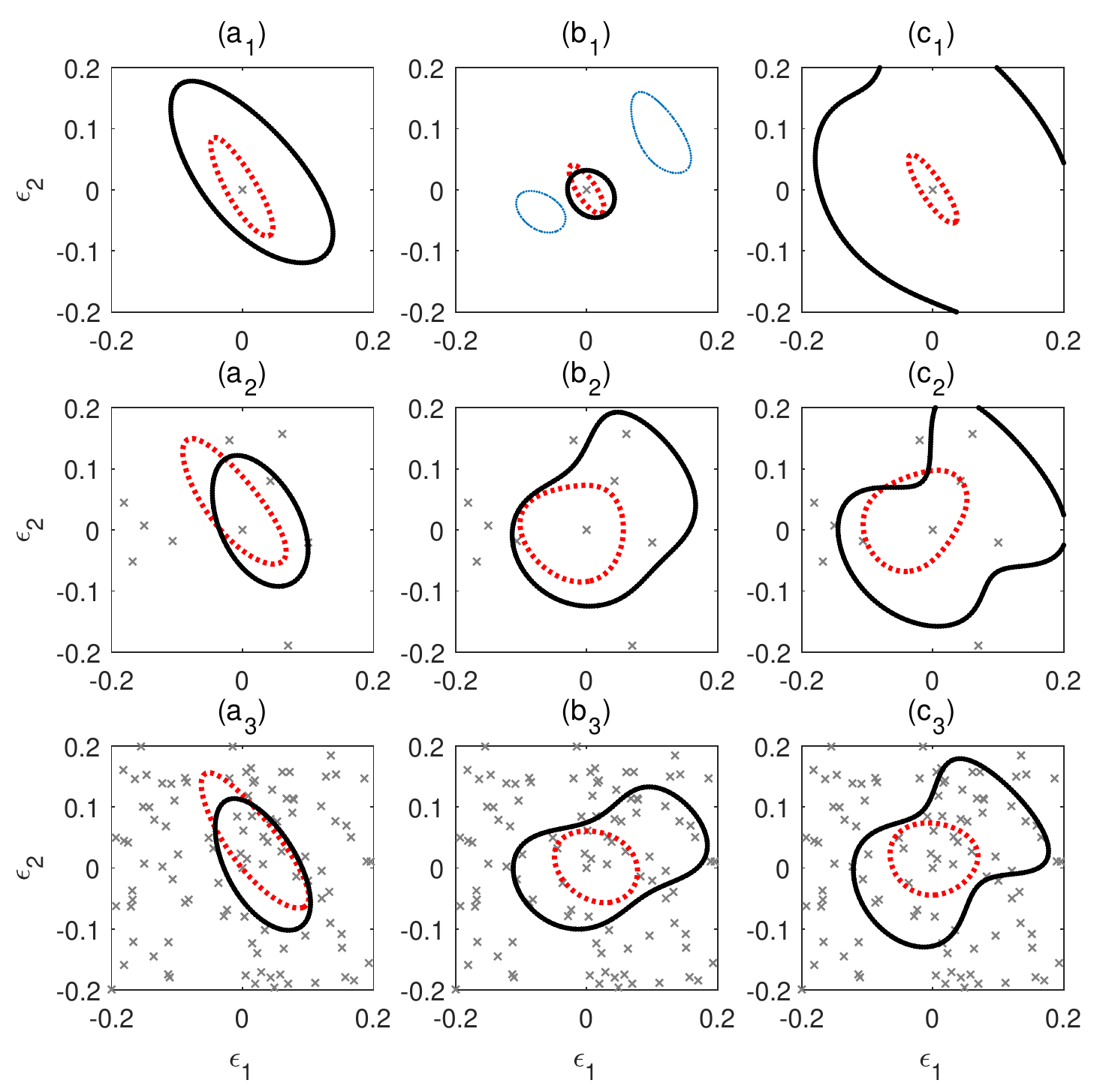}
\caption{The robustness (quantified by the area enclosed by the level set at $0.001$) of control fields obtained by b-GRAPE (black solid) and s-GRAPE (red dotted) algorithms. The time duration $T$ and number of control segments $M$ are, respectively, (a$_1$-a$_3$) $T=5$ and $M=50$; (b$_1$-b$_3$) $T=10$ and $M=50$; (c$_1$-c$_3$) $T=10$ and $M=100$. The corresponding batch sizes are $B=1$ (1st row), $B=10$ (2nd row) and $B=100$ (3rd row) }
\label{fig:robustness}
\end{figure}

We also test the performance of b-GRAPE with less available control resources, i.e., using shorter time duration ($T=5$) and less number ($M=50$) of control segments. The simulations are all stopped after evaluating one million samples. The simulations consistently approve the superiority of b-GRAPE (bold black) over s-GRAPE (dotted red) in all cases shown in Fig.~\ref{fig:robustness}. However, not surprisingly, the robustness is less enhanced when the control is more limited. In fact, b-GRAPE should be more advantageous under such circumstances because otherwise the search will be more easily trapped by local optima.

The simulations also includes an exceptional example shown in Fig.~\ref{fig:robustness}(b$_1$), whose robustness achieved by b-GRAPE is supposed to be stronger than those in Fig.~\ref{fig:robustness}(a$_1$) for the control time is longer and Figs.~\ref{fig:robustness}(b$_2$-b$_3$) for the the batch size is smaller. This poor solution results from an instable training process, during which the average infidelity over training batches rises up after about exploring about 800 thousands of samples. The level set at 0.001 (in blue) corresponding to the best solution before losing stability is also depicted in Fig.~\ref{fig:robustness}(b$_1$), which is disconnected due to the coexistence of two minima like in Fig.~\ref{fig:robust3D}. Note that the stochastic training process is not always instable, and very robust solutions can be obtained after restarting the stochastic b-GRAPE optimization, reselecting the initial guess or decrease the learning rates.

\subsection{Example 2: time-varying noise}
To demonstrate the applicability of the b-GRAPE algorithm to more general uncertainties, we consider the following single-qubit system that contains time-varying noises:
\begin{equation}\label{}
  H(t)=[1+n(t)]\cdot\left[u_x(t)\sigma_x+u_y(t)\sigma_y\right],
\end{equation}
where $\sigma_{x,y}$ are the Pauli matrices with $u_{x,y}(t)$ being the Rabi driving fields. The noise $n(t)$ represents the multiplicative time-varying noises in the control amplitudes.

Since the qubit is insensitive to high-frequency noises, we only sample the low-frequency noises as follows:
\begin{equation}\label{}
  n(t)=\sum_{k=1}^{10}\left(a_k\cos\omega_kt+b_k\sin\omega_kt\right),
\end{equation}
where the frequencies components $\omega_k$ are uniformly sampled from $0$ to $2\pi$rad/s and the amplitudes $a_k$ and $b_k$ are sampled from a Gaussian distribution $\mu(0,0.05)$. These parameters form a 30-dimensional sample space, to which the sampling-based algorithms can hardly handle because the a formidably large number samples will be required.

In the simulation, the target unitary transformation is chosen as the qubit flip, i.e., a $\pi$-rotation $U_f=R_x(\pi)$
around the $x$-axis. In absence of noises, the rotation can be easily achieved by applying an arbitrary $u_x(t)$ whose pulse area is $\pi$ (with $u_y(t)$ being turned off), e.g., rectangular or Gaussian. However, robustness is not guaranteed for these pulses.

\begin{figure}
\centering
\includegraphics[width=1\columnwidth]{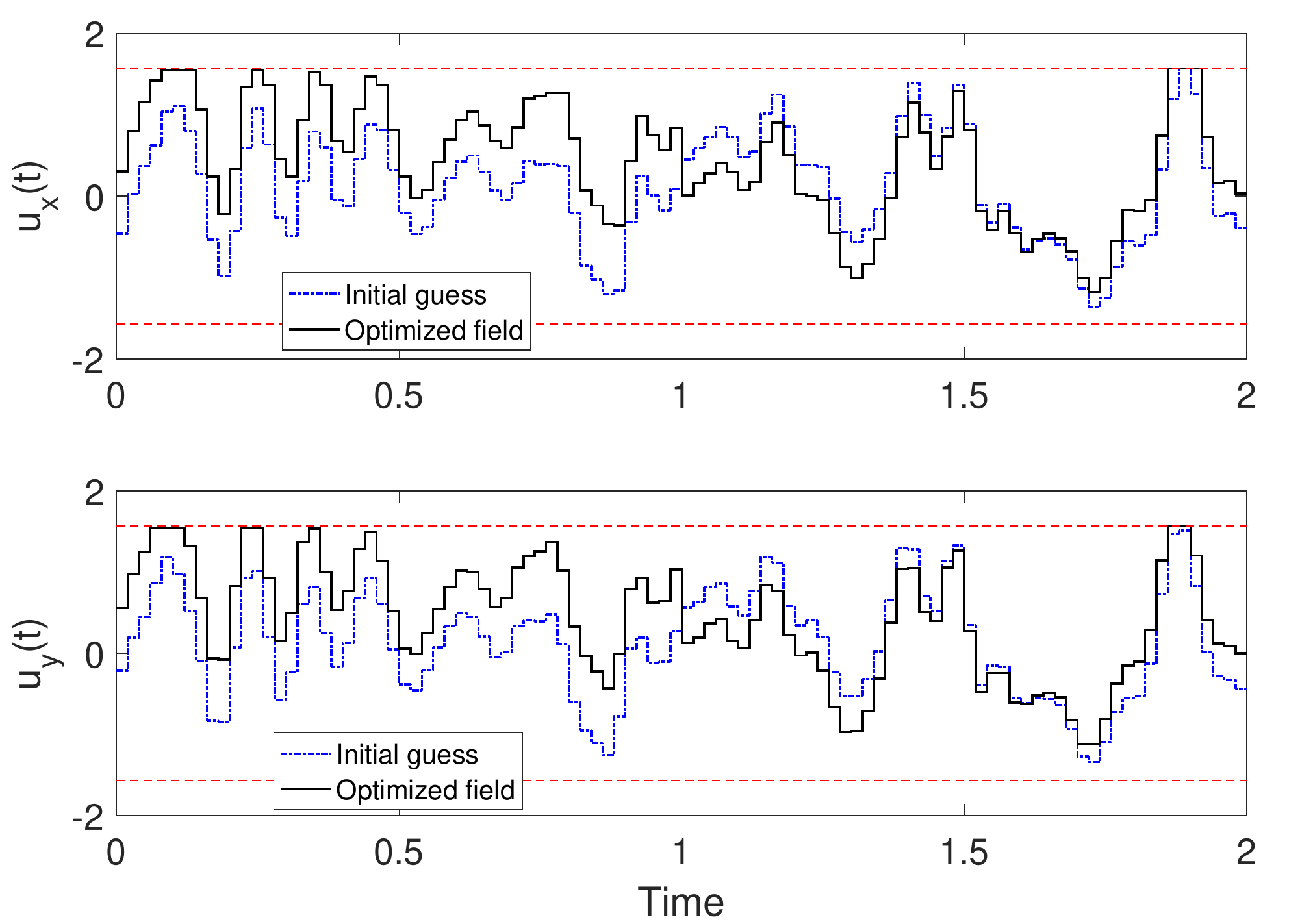}
\caption{The initial guess and optimized waveforms of the $x$-axis and $y$-axis control fields resulting fields.}
\label{fig:Field}
\end{figure}

We set the simulation time as $T=2$ and bound the control fields by $|u_{x,y}(t)|\leq\pi$. The batch size is chosen to be $B=10$ and totally 10000 iterations are performed, after which the average fidelity is reduced to be below $10^{-2}$. The waveforms of the initial guess and the optimized field are shown in Fig.~\ref{fig:Field}.

\begin{figure}
\centering
\includegraphics[width=1\columnwidth]{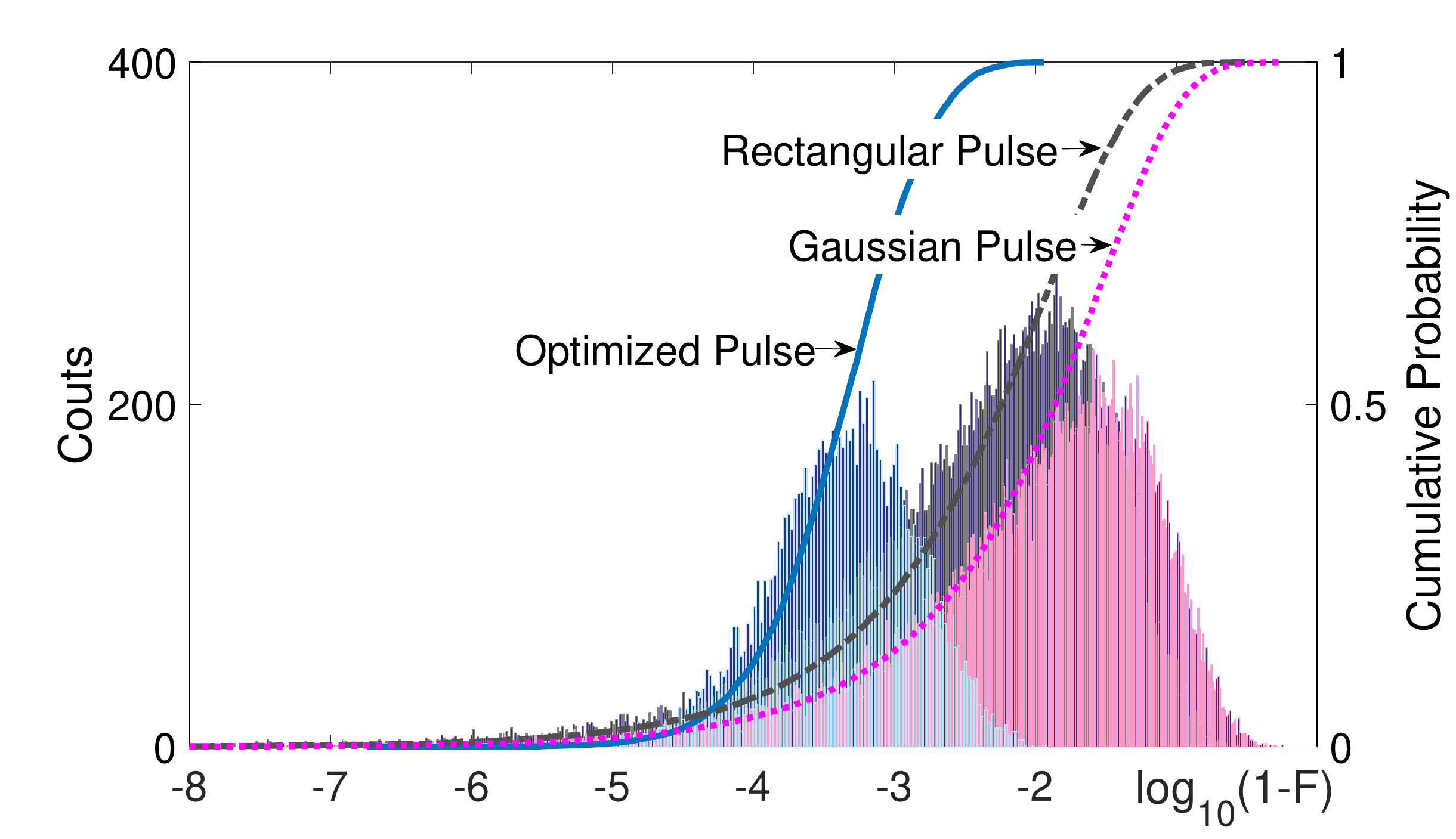}
\caption{Error distribution of single-qubit quantum controls counted from 10000 random samples of multiplicative noises in the amplitudes, where the target is the qubit flip transformation. The field optimized by b-GRAPE is very strong in that the probability for the error to be below $10^{-2}$ is almost $100\%$, which is much higher than those of rectangular and Gaussian $\pi$-pulses.}
\label{fig:Noise}
\end{figure}

To see how the robustness is enhanced, we test the obtained control field by analyzing the statistical distribution of the gate errors using 10000 random noise samples picked from the same distribution. As shown in Fig.~\ref{fig:Noise}, the gate error under the optimized control exhibits a typical Gaussian distribution whose center is below $10^{-3}$. We also depict the cumulative probability distribution function, from which it can be clearly seen that the probability for the error to be below $10^{-2}$ is almost $100\%$, and the probability for the error to be below $10^{-3}$ is about $76\%$. We also evaluate the robustness of the standard rectangular and Gaussian $\pi$-pulses, whose probabilities for the error to be below $10^{-2}$ are $62\%$ and $43\%$, respectively. Apparently, the optimized control fields is much more robust to the time-varying noises.

\section{Conclusion}\label{Sec:Conclusion}
 To conclude, we proposed b-GRAPE, a deep-learning illuminated algorithm, for efficiently discovering highly robust quantum controls in high-precision regime. The algorithm can be easily implemented by randomizing the renowned GRAPE algorithm with batches of samples, and numerical simulations demonstrate its effectiveness owing to the endowed ability of exploring uncountably many uncertainty samples, and the ability of escaping poor optima driven by the batch-induced randomness. Our algorithm can also be conveniently paralleled. Although the theoretically best performance is achieved when $B=1$, in practice we can adequately increase the batch size to improve the computational efficiency as well as the algorithmic stability.

In our numerical tests, no evident traps (i.e., local optima that are far from global optima) were encountered, which exhibit nice control landscapes that have been observed in both quantum control \cite{Rabitz2004,Rabitz2005,Wu2008b} and deep learning \cite{Kawaguchi2016,Zhou2017} studies. These results are closed related to the control landscape of uncertain quantum systems, and a further study on this topic will be very important to a full understanding of robust quantum control problems.

This work is only the start of a potentially large family of robust control design algorithms. As can be seen in our simulations, there is still much room for the control robustness to be enhanced, e.g., by fine tuning the learning rates, increasing the number of iterations and introducing more advanced tuning strategies. More DNN tuning skills developed in the DL practice can be easily transferred here, e.g., Adam, AdaGrad, REProp \cite{Bottou2018} or Newton-like stochastic gradient algorithms, namely Newton sketch \cite{Pilanci2017}), as well as gradient-free algorithms (e.g., genetic or differential evolution algorithms). We expect that such DL-inspired algorithms will produce signification impacts on the design of high-quality controls over quantum information processing hardware.



\end{document}